# $Mn_2V_{0.5}Co_{0.5}Z$ (Z= Ga, Al) Heusler alloys: Fully compensated ferrimagnets with high $T_c$ and compensation temperature

P V Midhunlal[1], J Arout Chelvane[2], D Prabhu[3], Raghavan Gopalan[3], and Harish Kumar N [1*]

1. Department of Physics, Indian Institute of Technology-Madras, Chennai-600036, India.
2. Defense Metallurgical Research Laboratory, Kanchanbagh (PO), Hyderabad-500058, India.
3. International Advanced Research Centre for Powder Metallurgy and New Materials (ARCI), Chennai-600113, India.

*nhk@iitm.ac.in

**Abstract:** High $T_C$ fully compensated ferrimagnets are potential candidates for spin transfer torque based spintronic devices. We report the structural and magnetic properties of high $T_C$ fully compensated ferrimagnets $Mn_2V_{0.5}Co_{0.5}Z$ (Z=Ga, Al) in the melt-spun ribbon and arc melted bulk form. While the parent alloys $Mn_2YZ$ (Y=V, Co; Z= Ga, Al) exhibits a magnetic moment value around 2 $\mu_B$/f.u., the $Mn_2V_{0.5}Co_{0.5}Ga$ alloy exhibits room temperature nearly fully compensated moment value of 0.09 and 0.13 $\mu_B$/f.u. in the bulk and ribbon form respectively. For $Mn_2V_{0.5}Co_{0.5}Al$ this turned out to be 0.04 and 0.08 $\mu_B$/f.u. In Contrast to the bulk sample's Néel P-type ferrimagnetic behaviour, ribbon samples exhibit Néel N-type ferrimagnetic characteristic with a high compensation temperature of 420 K for Z=Ga and 275 K for Z=Al. The observed $T_C$ values are more than 640 K for all samples. The differences in the magnetic properties of arc melted and melt-spun alloys indicates that even a slight variation in stoichiometry and sample preparation method can influence the physical properties of a compensated system.

## 1. Introduction

Half-metallic antiferromagnets (HMAFs) or more precisely half-metallic fully compensated ferrimagnets (HMFCFis) are a new class of magnetic materials which exhibits zero macroscopic moment with high spin polarization at the Fermi level. These materials differ from the conventional antiferromagnets in such a way that the different magnetic sub-lattices are chemically or crystallographically inequivalent and the compensation occurs in a wide range of temperature. Also, the characteristic temperature is the Curie temperature ($T_C$) and not the Néel temperature as in the case of conventional antiferromagnets [1]. This interesting class of materials are expected to be utilized as tips in Spin-polarized Scanning Tunneling Microscopes (SP-STM) and electrode material in Spin Torque Transfer (STT) based Magnetic Tunnel Junctions (MTJs) which would be the building blocks of future Magnetic Random Access Memory (MRAM)[2,3]. Among the various HMAFs which includes certain double perovskites, dilute magnetic semiconductors and superlattices [4,5], Heusler alloys have got much interest due to their magnetic moment tunability and high $T_C$. The Slater-Pauling relation which expresses the total magnetic moment ($M_t$) to the total number of valence electron per unit cell ($Z_t$) can be utilized to design HMFCFis by varying the valence electron number. As per the rule, $M_t = Z_t - 24$ for full-Heusler alloys (general formula $X_2YZ$, X & Y are transition metal and Z is a main group element) and $M_t = Z_t - 18$ for half-Heusler alloys (chemical formula XYZ), zero magnetic moment state is expected for full-Heusler alloys with 24 valence electrons and half-Heusler alloys with 18 valence electrons[6]. The *ab initio* studies carried out by I.Galanakis *et al.* have shown a way of achieving half-metallic fully compensated ferrimagnetism in $Mn_2VAl$ and $Mn_2VSi$ Heusler alloys by substituting Co at the Mn site[7]. Motivated by this interesting theoretical observation, researchers have carried out experimental investigation on the structural and magnetic properties of $(Mn_{2-x}Co_x)VZ$ (Z = Ga, Al) Heusler alloys[8,9]. Even though the compensation was achieved in these alloys, the $T_C$ has drastically come down from more than 700 K to below the room temperature as the compensation point approaches (x=1) which is a drawback as far as the applications are concerned. At the same time, our recent investigation on the Co substitution at the V site of $Mn_2VZ$ (Z= Ga, Al) alloys could sort out the issue of decreasing $T_C$. $Mn_2V_{1-x}Co_xZ$ (Z = Ga, Al) alloys exhibited near zero moment state when x=0.5 by preserving high $T_C$ (more than 650 K)[10]. Here in this report we investigate the structural and magnetic properties of high $T_C$ fully compensated ferrimagnetic $Mn_2V_{0.5}Co_{0.5}Z$ (Z = Ga, Al) ribbons with high compensation temperature. The properties are compared with that of the bulk samples and the interesting deviations in the compensation behaviour are discussed.

## 2. Experimental details

$Mn_2V_{0.5}Co_{0.5}Z$ (Z = Ga, Al) bulk samples were prepared by arc melting the individual elements with high purity. Initially, titanium was melted several times to absorb any leftover oxygen in the chamber. Samples were melted several times after flipping to ensure homogeneity. Samples were sealed in an evacuated quartz tube for annealing. $Mn_2V_{0.5}Co_{0.5}Ga$ was annealed at 1073 K and $Mn_2V_{0.5}Co_{0.5}Al$ was annealed at 673 K for three days followed by furnace cooling. $Mn_2V_{0.5}Co_{0.5}Z$ (Z = Ga, Al) ribbon samples were prepared from the arc melted ingots by vacuum induction melting followed by melt spinning method. The molten alloys were ejected over rotating copper wheel rotating at 1000 rpm. Flakes with 30 - 40 µm thickness and approximate length of 10 mm and breadth of 2 mm were obtained. The structural characterization was carried out using Rigaku SmartLab high-resolution X-ray diffractometer with Cu-K$_\alpha$ radiation. The compositions and microscopic images of the ribbons were recorded using FEI-InspectF Scanning Electron Microscope (SEM). The magnetic measurements in the temperature range 5- 300 K were carried out using Quantum Design MPMS 3 SQUID VSM and high-temperature magnetic measurements (300–850 K) were carried out using Microsense Vibrating Sample Magnetometer, Model EZ9.

## 3. Results & Discussion

### 3.1 Structural properties

Fig. 1(a) - (f) shows the cross-sectional and surface SEM images of the ribbon samples. Surfaces morphology appeared to be different for the two sides for both the alloys (one side is smooth and the other side is rough). This is expected as one of the surfaces touches the wheel and the other surface exposed to the inert atmosphere during the melt spinning technique.

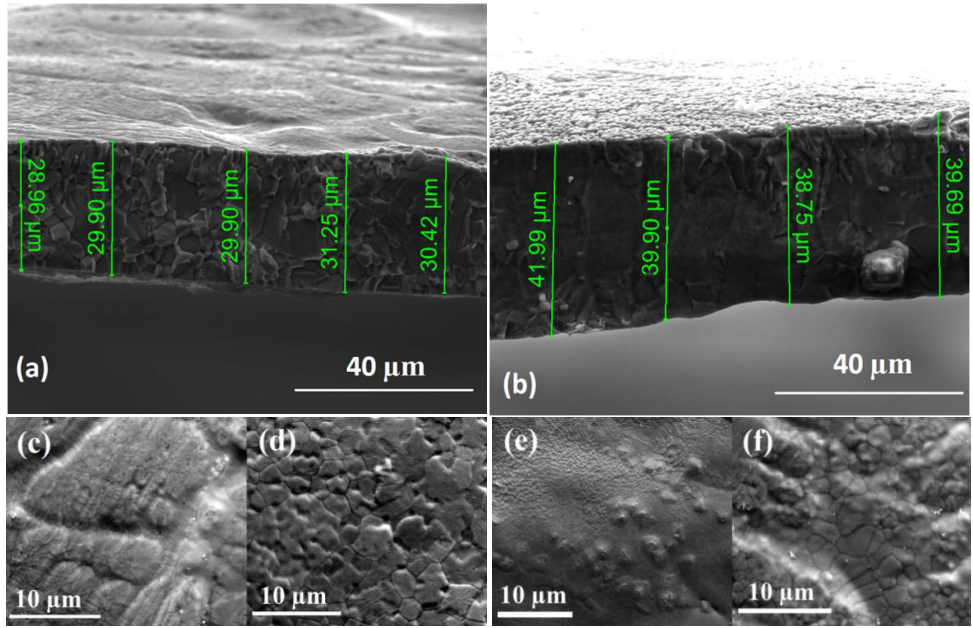

**Fig. 1(a) Cross-sectional SEM images of $Mn_2V_{0.5}Co_{0.5}Ga$ and (b) $Mn_2V_{0.5}Co_{0.5}Al$ ribbons. (c) & (d) surface images (both sides) of $Mn_2V_{0.5}Co_{0.5}Ga$ and (e) & (f) $Mn_2V_{0.5}Co_{0.5}Al$ ribbons.**

Fig. 2(a) and (b) shows the X-ray diffraction (XRD) patterns of $Mn_2V_{0.5}Co_{0.5}Z$ ribbon and bulk samples. There was not much difference in the XRD pattern of bulk and corresponding ribbon samples. The (111) and (200) Superlattice reflections were absent in the case of Z=Ga samples while Z=Al samples exhibited these reflections with low intensity. The absence/low intensity could be due to the similar atomic scattering factors of x-ray for the atoms which are in the same period in the periodic table. Here it is to be noted that as per the earlier reports, the parent $Mn_2VGa$ alloy has not exhibited any superlattice reflection in the XRD pattern and the neutron diffraction pattern has shown the superlattice peaks with huge intensity[11,12]. It is reported that $Mn_2VZ$ alloys crystallize in the cubic $L2_1$ structure (space group:225, $Fm\bar{3}m$) and $Mn_2CoZ$ alloys crystallize in the cubic $X_a$ structure (space group:216, $F\bar{4}3m$)[12–14]. As the parent alloys/end members possess different crystal structure, it is difficult to predict the crystal structure of $Mn_2V_{0.5}Co_{0.5}Z$ samples. Bearing this in mind, Rietveld refinement of the XRD patterns were carried out by using FullProf software[15]. assuming both $L2_1$ and $X_a$ structures for all samples. Interestingly both the structures have given a similar fit with nearly the same lattice parameter and $\chi^2$ values. To have a better understanding, the XRD patterns were simulated for both the structures and it was evident that the patterns are indistinguishable. Since the lattice parameters are same for both the structures, refined pattern assuming $X_a$ structure is shown in Fig. 2 (a) and (b). The obtained lattice parameters are 5.872 and 5.857 Å for Z=Ga and Al ribbon respectively which are close to the bulk values. The estimated composition, lattice parameter and other magnetic parameters are shown in table I.

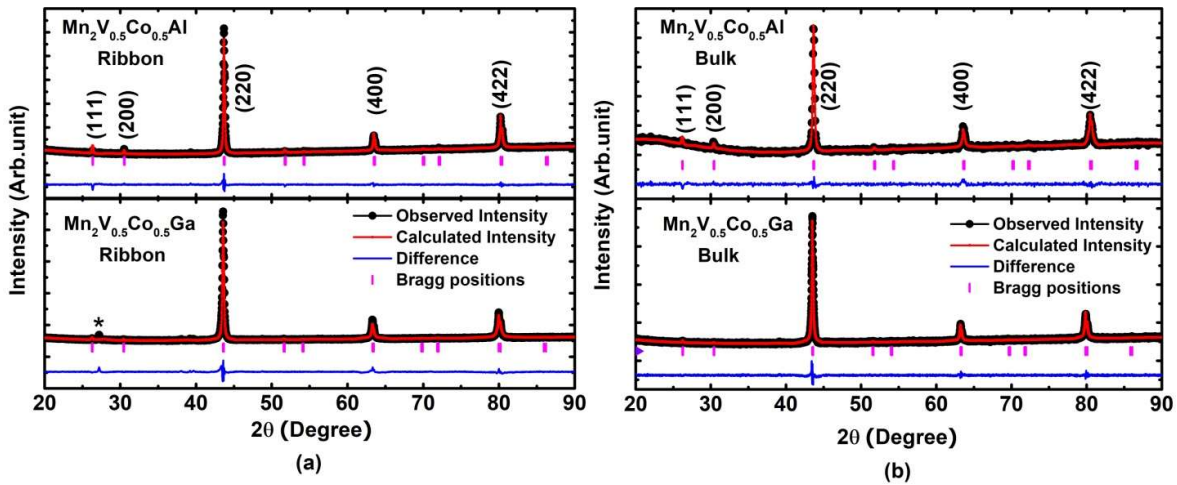

Fig. 2(a) Refined X-ray diffraction patterns of $Mn_2V_{0.5}Co_{0.5}Z$ ribbon samples and (b) Bulk samples.

| Alloy | Type | EDS Composition | Lattice parameter (Å) | M ($\mu_B$/f.u.) 5 K | M ($\mu_B$/f.u.) 300 K | $T_C$ (K) |
|---|---|---|---|---|---|---|
| $Mn_2V_{0.5}Co_{0.5}Ga$ | Ribbon | $Mn_{2.09}V_{0.51}Co_{0.52}Ga_{0.87}$ | 5.872 | 0.23 | 0.09 | 672 |
| | Bulk | $Mn_{2.08}V_{0.52}Co_{0.47}Ga_{0.91}$ | 5.878 | 0.10 | 0.13 | 706 |
| $Mn_2V_{0.5}Co_{0.5}Al$ | Ribbon | $Mn_{2.03}V_{0.51}Co_{0.52}Al_{0.91}$ | 5.857 | 0.10 | 0.04 | 641 |
| | Bulk | $Mn_{2.02}V_{0.48}Co_{0.50}Al_{1.00}$ | 5.825 | 0.06 | 0.08 | 659 |

**Table I: Structural and magnetic parameters of $Mn_2V_{0.5}Co_{0.5}Z$ alloys in the ribbon and bulk forms**

## 3.2 Magnetic properties

Our earlier report has investigated the detailed magnetic properties of the $Mn_2V_{1-x}Co_xZ$ (Z = Ga, Al) bulk alloys. The obtained moment values for the end members in the series are 1.80, 2.05, 1.83 and 2.06 $\mu_B$/f.u. for $Mn_2VGa$, $Mn_2CoGa$, $Mn_2VAl$ and $Mn_2CoAl$ respectively[10]. Here we focus on the x=0.5 member in the bulk and ribbon forms, which exhibits total magnetic moment compensation with high $T_C$. The isothermal magnetization curves at 5 K and 300 K for the ribbons and bulk samples with x=0.5 are shown in Fig. 3 (a)-(d). The nearly compensated magnetic moment values obtained through Honda plots of M-H curves at 5 K are 0.23 and 0.10 $\mu_B$/f.u. for Z= Ga and Al ribbon respectively. This is slightly higher than the magnetization values of the corresponding bulk samples which are 0.1 and 0.06 $\mu_B$/f.u. The magnetic parameters are shown in table I. The moment was found to decrease with increase in temperature for the ribbon samples and an increment was observed for the bulk samples as shown in Fig. 3. Moment values at 300 K are 0.09 and 0.04 $\mu_B$/f.u. for Z= Ga and Al ribbons respectively. This indicates that even though magnetic moment compensation happens in both ribbon and bulk samples, the effect of temperature on the magnetic properties are different. This is more evident from the temperature variation of magnetization (M-T) measured in the range 5- 850 K as shown in Fig. 4 (a)-(d). Since the low-temperature M-T curves (5-300 K) and high-temperature M-T curves (300-850 K) are recorded using different instruments, combined normalized curves are shown in the figures.

The low temperature M-T curves measured at 100 Oe field and in the range 5-300 K are also shown in the insets of Fig. 4 (a)-(d).

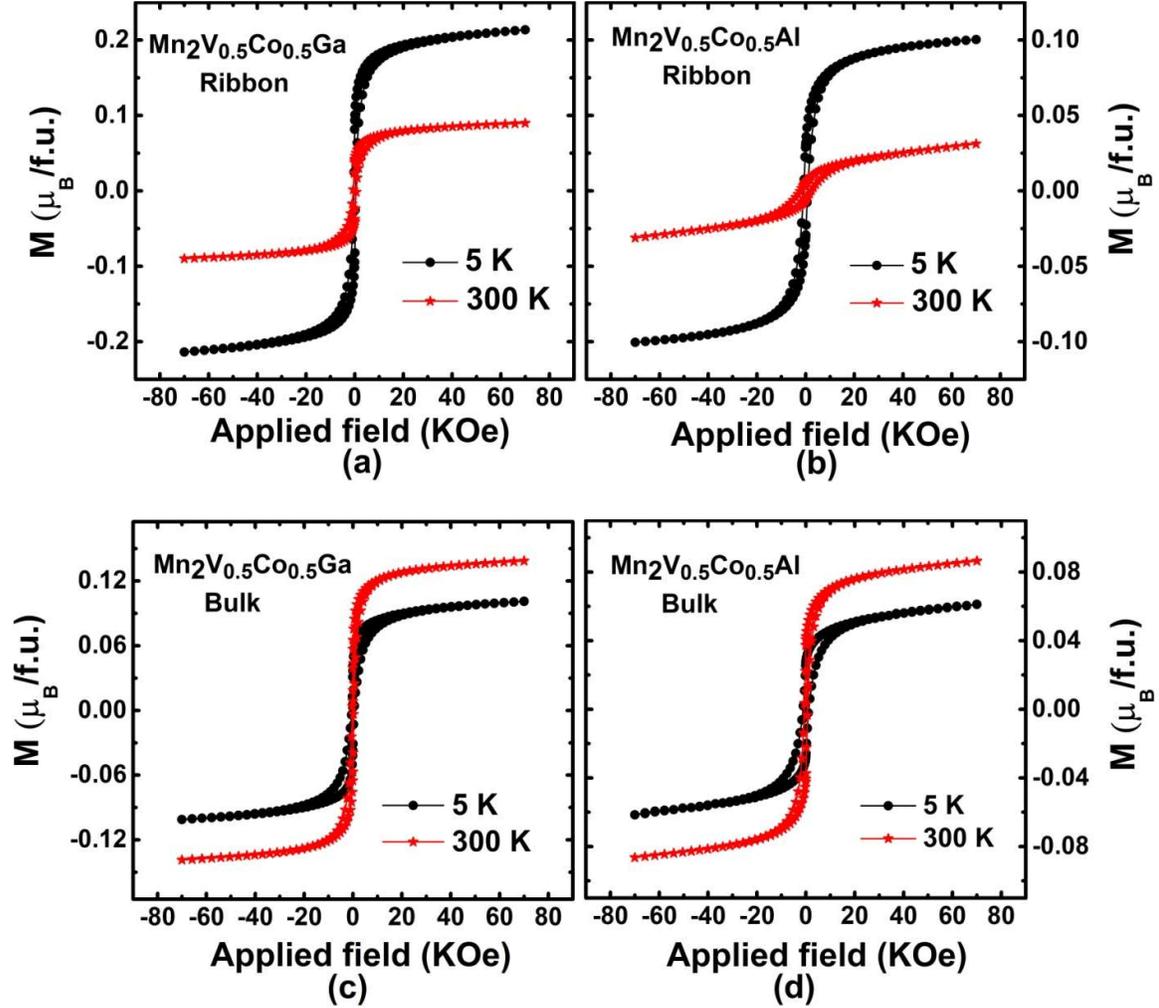

**Fig. 3(a) isothermal magnetization curves of $Mn_2V_{0.5}Co_{0.5}Ga$ and (b) $Mn_2V_{0.5}Co_{0.5}Al$ ribbons measured at 5 and 300 K. (c) & (d) isothermal magnetization curves of $Mn_2V_{0.5}Co_{0.5}Ga$ and $Mn_2V_{0.5}Co_{0.5}Al$ bulk samples.**

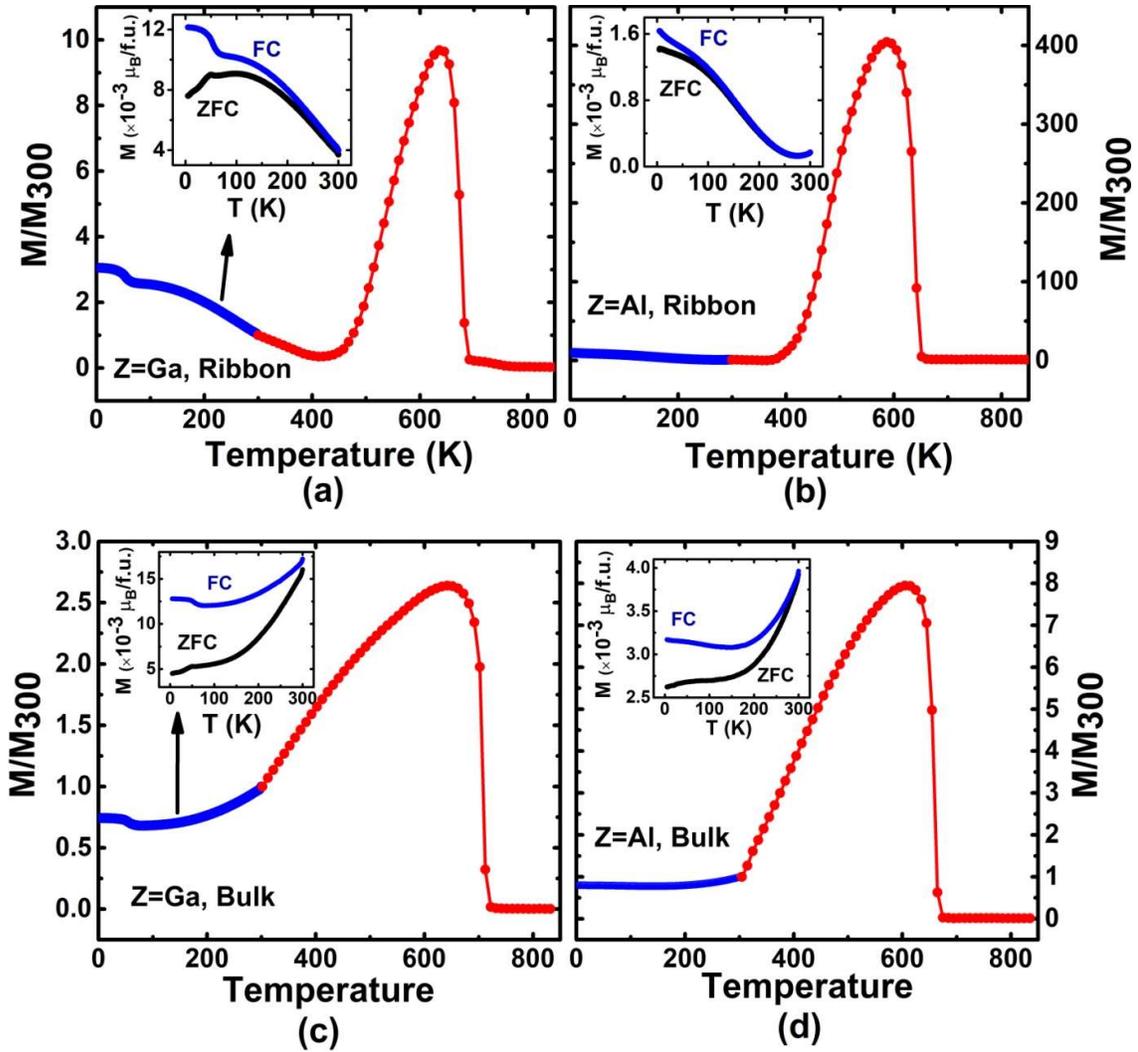

**Fig. 4(a)** M-T curves of $Mn_2V_{0.5}Co_{0.5}Ga$ and **(b)** $Mn_2V_{0.5}Co_{0.5}Al$ ribbons measured at 100 Oe field **(c) & (d)** M-T curves of $Mn_2V_{0.5}Co_{0.5}Ga$ and $Mn_2V_{0.5}Co_{0.5}Al$ bulk samples. Insets shows the low temperature ZFC and FC

The M-T curves for the ribbons show different behaviour compared to the bulk samples. For the ribbons, magnetic moment was found to decrease with increase in temperature from a non-zero value and nearly full compensation occurs around 420 K for Z=Ga and 275 K for Z=Al. This behaviour is in agreement with the decrease in moment observed in the M-H curves recorded at different temperatures. By increasing the temperature further, ferrimagnetic to paramagnetic transition has been occurred and the $T_C$ values were found to be 672 K and 641 K for Z=Ga and Z= Al ribbons respectively. This is slightly less compared to the bulk values of 706 K and 659 K for Z=Ga and Z=Al respectively. The M-T curves clearly show the presence of a full compensation temperature which was absent in the case of bulk samples. A similar behaviour is observed in the compensated ferrimagnet $Mn_{1.5}FeV_{0.5}Al$ (bulk arc melted sample)[16,17]. It is to be noted that the compensation temperature for $Mn_{1.5}FeV_{0.5}Al$ bulk alloy was 127 K (which has been tuned up to 226 K by varying the stoichiometry) which is much lower than the compensation temperature of $Mn_2V_{0.5}Co_{0.5}Ga$ ribbon reported in this paper (420 K). In the earlier reported $Mn_{1.5}FeV_{0.5}Al$ system, three different kinds of M-T behaviour was observed, when the stoichiometry was varied. The first case is the fully compensated ferrimagnetic state where a zero moment is observed near 0 K and it is maintained up to a certain temperature (50 K in the case of $Mn_{1.5}FeV_{0.5}Al$) followed by an increase in magnetization and then transition from ferrimagnetic state to paramagnetic state. The authors could also simulate this M-T curve using a molecular field model for a two sub-lattice ferrimagnet. The second case is the overcompensated ferrimagnetic state where the compensation temperature was shifted to a certain temperature (308 K) by choosing an appropriate stoichiometry for the parent alloy. Here the magnetization is non-zero near 0 K and then decreases with increase in temperature, reaching a fully compensated state and then follow similar behaviour as in the first case. This is Néel N-type ferrimagnetic behaviour. In the third case which is known as the Néel P-type ferrimagnetic behaviour, full compensation does not occur in the entire temperature range and the moment keeps on increasing with the increase in temperature[18]. Now comparing these three M-T characteristics, it is clear that the $Mn_2V_{0.5}Co_{0.5}Z$ ribbon samples fall in the second category which is Néel N-type ferrimagnet. But the decrease in the magnetization up to the full compensation temperature is not as fast as that of $Mn_{1.5}FeV_{0.5}Al$ alloy indicating different exchange coupling strength for the different magnetic sublattices. As far as the bulk samples are concerned, a full compensation point was not observed as in the case of ribbon samples. An increase in the magnetic moment was observed till the $T_C$ indicating that the bulk samples could be Néel P-type ferrimagnets. The $T_C$ was found to be 706 K and 659 K for Z=Ga and Al respectively. It is to be noted that even though the moment is varying with the temperature, the magnitudes show that the samples are in almost a near compensated state at least up to 420 K for the ribbon and 300 K for the bulk samples. It is expected that a slight variation in the composition could highly affect the magnetic sub-lattices ordering of a fully compensated system which would give a different magnetic response with the temperature as in the case of $Mn_{1.5}FeV_{0.5}Al$ [17]. In addition to this, the sudden quenching of molten liquid would have affected the magnetic sub-lattices ordering of ribbon samples (quenching rate is around $10^4$ K/s). $Mn_2V_{0.5}Co_{0.5}Ga$ ribbon and bulk samples exhibit a negligibly small transition around 50 K. Unlike the Z=Ga samples, Z=Al samples (ribbon and bulk) has not shown any transition in the low-temperature regime indicating that the presence of small second magnetic phase is not influencing the temperature dependent compensation in Z=Ga sample.

## 4 Conclusions

The magnetic properties of fully compensated ferrimagnets $Mn_2V_{0.5}Co_{0.5}Z$ (Z=Ga, Al) show distinctly different magnetic properties in their ribbon and bulk form. While the ribbon samples exhibit Néel N-type ferrimagnetic behaviour with high compensation temperature, the bulk samples exhibit Néel P-type ferrimagnetic behaviour without any full compensation temperature. Even though there exists a temperature dependent moment variation in the samples, an overall nearly fully compensated state is preserved at least up to 420 K for the ribbons and 300 K for the bulk samples. These materials having zero moment, high $T_C$ and wide temperature range of compensation would be attractive for the future spintronic devices utilizing fully compensated ferrimagnets.